\documentclass[conference]{IEEEtran}
\IEEEoverridecommandlockouts
\usepackage{url}

\usepackage{cite}
\usepackage{amsmath,amssymb,amsfonts}
\usepackage{algorithmic}
\usepackage{graphicx}
\usepackage{textcomp}
\usepackage{xcolor}
\usepackage{subcaption}
\def\BibTeX{{\rm B\kern-.05em{\sc i\kern-.025em b}\kern-.08em
    T\kern-.1667em\lower.7ex\hbox{E}\kern-.125emX}}
    
\begin{document}

\title{Sensing-Based Communication: A Non-Orthogonal Passive Reflection Framework for ISAC}

\author{\IEEEauthorblockN{Xin Xiong\IEEEauthorrefmark{1}, Mathew Salas\IEEEauthorrefmark{1}, Sihua Shao\IEEEauthorrefmark{1}, Chen Chen\IEEEauthorrefmark{2}, Mostafa Hassanalian\IEEEauthorrefmark{3} and Rafael Fierro\IEEEauthorrefmark{4}}
\IEEEauthorblockA{\IEEEauthorrefmark{1}Department of Electrical Engineering,
Colorado School of Mines, Golden, Colorado 80401, USA
\\\IEEEauthorrefmark{2}School of Microelectronics and Communication Engineering, Chongqing
University, Chongqing 400044, China
\\\IEEEauthorrefmark{3}Department of Mechanical Engineering, New Mexico Tech, Socorro, New Mexico 87801, USA 
\\\IEEEauthorrefmark{4}Department of Electrical and Computer Engineering, University of New Mexico, Albuquerque, New Mexico 87131, USA\\
Emails: xin\_xiong@mines.edu, mathew\_salas@mines.edu, sihua.shao@mines.edu, c.chen@cqu.edu.cn, \\mostafa.hassanalian@nmt.edu, rfierro@unm.edu
}
\vspace{-30pt}

\thanks{This work was supported by the National Science Foundation under Award no. CNS-2430772 and CNS-2431272. (\textit{Corresponding author: Sihua Shao.})}
}

\maketitle

\begin{abstract}
While integrating uplink without disrupting the primary downlink remains a critical challenge for existing reflective communication systems, this paper proposes a non-orthogonal passive integrated sensing and communication (ISAC) mechanism termed Sensing-based Communication. Originating from and validated within an optical retro-reflective system, this mechanism possesses an underlying wave-agnostic architecture with the potential to be generalized across other physical modalities. It leverages incident downlink signals for parasitic modulation via passive tags. By relying solely on fundamental on-off keying (OOK) reflection, the tag unifies the reflective sensing channel and the uplink communication channel, and simultaneously achieves both sensing and duplex communication. Consequently, our approach introduces uplink capability without sacrificing the downlink data rate or consuming additional spectrum resources. At the receiver, we design a joint decoding architecture combining cross-correlation computation and successive interference cancelation (SIC) to fully exploit the degrees of freedom in broadband signals for multi-user interference suppression. We derive closed-form expressions for the interference and bit error rate (BER) assuming zero-mean Gaussian signals. Using an experimental testbed, we validate the pivotal theoretical finding, the scaling relationship of SINR with signal bandwidth and cross-correlation time—where linear regression yields a strong goodness-of-fit with $R^2 = 0.9688$. Furthermore, evaluations under noise-like orthogonal frequency division multiplexing (OFDM) waveforms exhibit high consistency between theoretical derivations and simulation data ($R^2 = 0.9372$, mean absolute percentage error, $\text{MAPE} = 14\%$), confirming the model's accuracy and waveform generalizability. Simulations validate our design, confirming zero-overhead passive communication. 
\end{abstract}


\section{Introduction}
While integrated sensing and communication (ISAC) is a vital trend for next-generation networks\cite{wang2023road}, in existing backscatter-ISAC literature, sensing and communication typically target distinct entities\cite{9737357, 10163897}. The work in \cite{sun2024study} presents a representative paradigm where the base station steers part of its signal toward passive tags for internet of things (IoT) communication, while relying on distinct echoes from separate, non-cooperative targets for environmental sensing. As reflective tags inherently evolve toward minimalist architectures and ultra-low power consumption\cite{10130082}, how to seamlessly unify both sensing and communication capabilities over a passive tag without hardware overhead remains a challenge. 

Within acoustic reflective systems, existing passive tags struggle more to simultaneously support full-duplex communication and reflective sensing. The acoustic passive tags typically exhibit a narrow bandwidth limited to a few kilohertz (kHz)\cite{jang2019underwater} and at most 20 kHz\cite{ghaffarivardavagh2020ultra}. Despite this severe constraint, frequency division multiplexing (FDM) remains the predominant multi-access strategy. For example, the system in \cite{ghaffarivardavagh2020ultra} utilizes a single-frequency sinusoidal signal for the downlink, while tags perform out-of-band modulation at different switching rates to distribute their uplink signals across distinct frequency bands. \cite{wang2025mobile} employs modulated chirp signals to facilitate very limited downlink responses. A critical drawback uniting these approaches is that supporting either uplink transmission or reflective localization comes at the direct expense of primary downlink capacity. A pure single-frequency sinusoidal signal lacks any downlink data transmission capability. In the system presented in~\cite{10801869}, since all time and spectral slots are occupied for localization, neither legitimate downlink nor uplink communication data can be conveyed. The bottleneck of acoustic transmission is the extreme scarcity of frequency resources of the reflective tags, posing a challenge to support both communication and sensing tasks.

This challenge is also prevalent in optical retro-reflective ISAC (RO-ISAC) systems\cite{11695581}. Conventional architectures require active light emitters and time- or wavelength-division multiplexing\cite{wang2025bidirectional,chen2025full} to separate echoes from active uplinks. Because the reflective link is parasitic on the downlink carrier, to isolate this reflective echo from active uplink communications, the system must partition resources in time or frequency, mimicking traditional frequency-division duplexing (FDD) or time-division duplexing (TDD) schemes while enforcing orthogonal time-division multiplexing (TDM) or FDM among uplink tags. However, the narrow bandwidth of passive components renders such orthogonal multiplexing inefficient. To overcome this, our prior work\cite{11298502,11585627 } introduced a non-orthogonal reflection mechanism that successfully enabled multi-tag simultaneous access and high ranging accuracy. Building upon this foundation, this paper advocates an \textit{asymmetric system architecture} to minimize tag complexity and power consumption. By delegating all heavy tasks—multi-user separation, interference cancellation, and decoding—to the receiver, tags only perform primitive baseband on-off keying (OOK) to directly embed uplink data into the reflected channel. This paradigm unifies communication and sensing without compromising downlink capacity, fitting ultra-low-power passive IoT tags.

Motivated by this asymmetric design philosophy, our main contributions are summarized as follows:

\begin{itemize}
    \item Novel Asymmetric ISAC Mechanism: We propose a non-orthogonal sensing-based communication framework that leverages downlink signal parasitic modulation. It achieves multi-user concurrent access with zero-resource-occupancy on the primary downlink, maintaining extreme hardware simplicity at the tags.
    
    \item Experimental Validation via RO-ISAC Platform: We evaluate the mechanism on an experimental RO-ISAC platform. Our empirical results corroborate the cornerstone of our theoretical analysis, confirming the derived relationship between signal-to-interference-plus-noise ratio (SINR), bandwidth, and correlation time.
    
    \item Theoretical Performance Bounds and System Simulation: We establish closed-form expressions for the interference, noise, and bit error rate (BER) based on a zero-mean Gaussian model. Furthermore, exploiting the Gaussian-like characteristics of orthogonal frequency division multiplexing (OFDM) waveforms, we extend the theoretical framework to concurrent multi-user environments, with simulation results corroborating our analytical derivations and confirming overall system feasibility.
\end{itemize}

\section{System Model}
\begin{figure}[t]
    \centering
    \includegraphics[width=\linewidth]{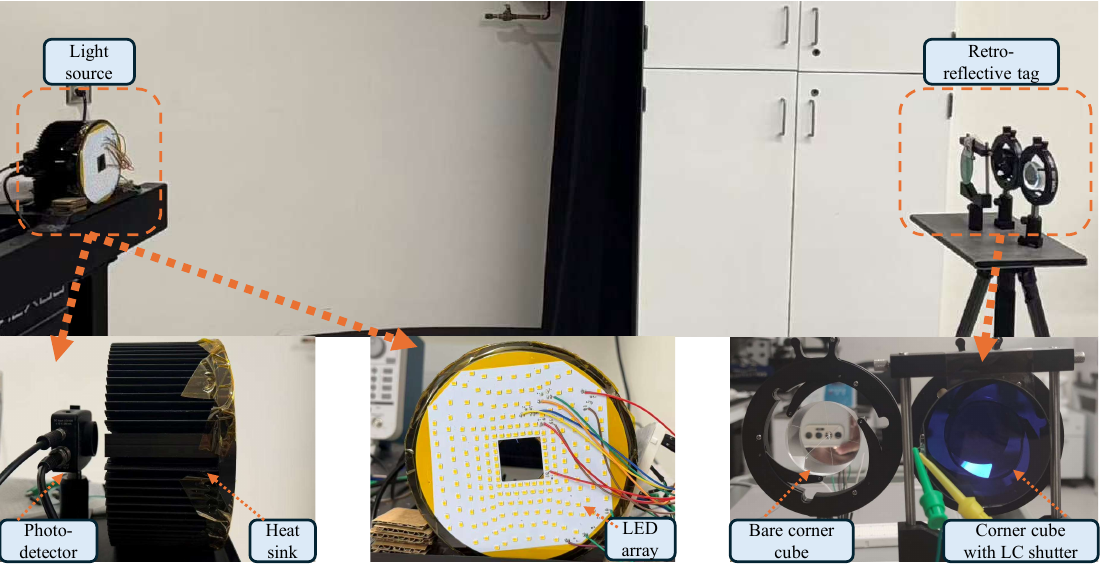}
    \caption{Experimental setup of the RO-ISAC system.}
    \label{fig:ro_isac_platform}
\vspace{-6pt}
\end{figure}

\begin{figure*}[t]
    \centering
    \includegraphics[width=\linewidth]{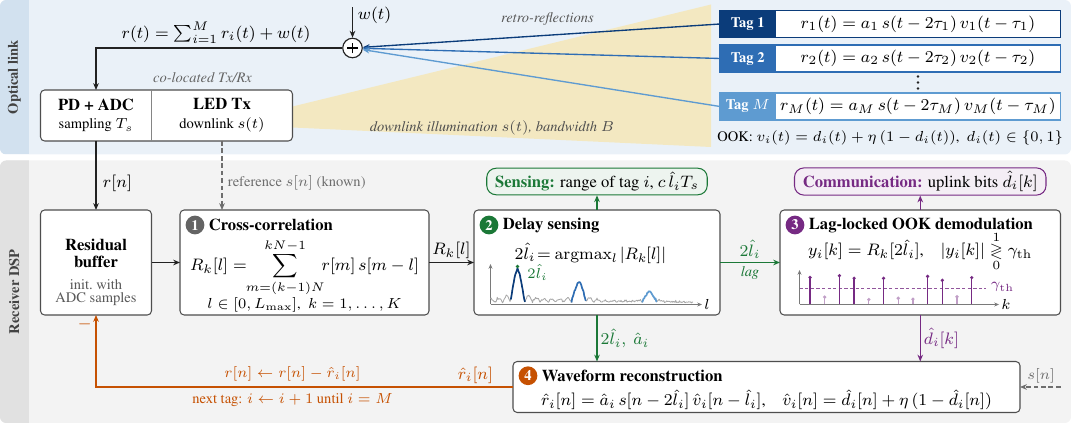}
    \caption{Architecture of the signal processing workflow.}
    \label{fig:sic_flowchart}
\vspace{-6pt}
\end{figure*}

Our derivations are formulated upon the RO-ISAC architecture depicted in Fig.~\ref{fig:ro_isac_platform}. The setup comprises an LED array source with heat sinking, retro-reflective tags, and a photodetector (PD) receiver. Each tag combines bare and liquid-crystal (LC) shutter-modulated corner cubes to achieve tunable modulation depth \cite{11298502}. A convex lens is positioned to focus the retro-reflected light onto the PD positioned behind it. The received optical signal is then digitized for baseband processing. The signal processing pipeline, from multi-tag superposition to sensing-based demodulation, is summarized in Fig.~\ref{fig:sic_flowchart}.

\subsection{Downlink Transmission}
Let the primary downlink signal $s(t)$ be a band-limited Gaussian signal with bandwidth $B$, which effectively approximates wideband signals like OFDM. Based on the retro-reflective optical channel model \cite{shao2023r}, the received optical power for tag $i$ follows the free-space path-loss formulation. Under direct detection with a photodiode responsivity $R$ (normalized to unity, i.e., $R = 1\,\mathrm{A/W}$, without loss of generality), the baseband electrical amplitude attenuation coefficient $a_i$ scales with the received optical power as
\begin{equation}
    a_i = R \frac{P_{r,i}}{P_t} = \frac{(m_l+1)A_{\mathrm{PD}}}{8\pi d_i^2} \cos^{m_l+1}(\theta_i) \epsilon_i,
    \label{eq:attenuation_coeff}
\end{equation}
where $P_t$ and $P_{r,i}$ denote the transmitted and received optical powers, $m_l$ is the Lambertian index, $A_{\mathrm{PD}}$ is the PD effective sensing area, $d_i$ is the distance to tag $i$, $\theta_i$ is the incidence angle, and $\epsilon_i$ represents the effective reflection area. This attenuation directly scales $s(t)$ at the receiver.

\subsection{Tag Parasitic Modulation}
The tags have extreme hardware simplicity, relying solely on ultra-low-power OOK modulation to retro-reflect the incident continuous-time downlink signal $s(t)$. In a multi-tag system comprising $M$ tags, the retro-reflection from the $i$-th tag introduces a specific time delay $\tau_i$ and an amplitude attenuation factor $a_i$, which corresponds to its two-way channel path loss. 

We introduce the \textit{modulation depth} of the tag, which essentially dictates the ratio of the reflected light amplitude in the ``OFF'' state to that in the ``ON'' state. We denote this residual amplitude fraction as $\eta$ (e.g., $\eta = 0.1$). In our hybrid design, this modulation depth $\eta$ is jointly determined by two distinct factors: (1) the light transmittance ratio between the ON and OFF states of the LC shutters; and (2) the structural reflectivity ratio between the \textit{unmodulated corner cubes} (bare corner cubes without LC shutters) and the \textit{modulated corner cubes} (those covered by LC shutters). 

Let $d_i(t) \in \{0, 1\}$ represent the binary OOK data sequence of the $i$-th tag. The modulation waveform $v_i(t)$ applied by the tag can be modeled as:
\begin{equation}
v_i(t) = d_i(t) + \eta \left( 1 - d_i(t) \right).
\end{equation}
Denoting the one-way propagation delay as $\tau_i$ (the two-way time delay is $2\tau_i$) and incorporating the amplitude attenuation factor $a_i$, the received reflected signal $r_i(t)$ from the $i$-th tag can be formulated as:
\begin{equation}
r_i(t) = a_i s(t - 2\tau_i) v_i(t - \tau_i).
\end{equation}

The received superimposed signal at the receiver comprises the reflections from the $M$ tags and noise. The received signal $r(t)$ is expressed as:
\begin{equation}
r(t) = \sum_{i=1}^{M} a_i s(t - 2\tau_i) v_i(t - \tau_i) + w(t),
\end{equation}
where $w(t)$ represents the additive white Gaussian noise (AWGN).

\section{Sensing-Based Non-Orthogonal Communication}

\subsection{Sensing-Based Communication Mechanism}
At the receiver, the signal is converted into the digital domain with a sampling period $T_s$. Let $n$ denote the discrete time index, such that $t = nT_s$. The continuous one-way delay $\tau_i$ is mapped to a discrete sample delay $l_i = \tau_i / T_s$. The discrete-time received signal is thus given by:
\begin{equation}
r[n] = \sum_{i=1}^{M} a_i s[n - 2l_i] v_i[n - l_i] + w[n].
\end{equation}

Tags $1$ to $M$ are sorted by increasing distance from the source, with adjacent spacing exceeding half the mainlobe width to prevent main-lobe interference. The receiver decodes signals successively, starting from the strongest signal which is tag $1$. The search lag is capped by the maximum delay $L_{\max}$ ($\text{max\_delay\_sample}$). To avoid state transitions during integration, the correlation window $N$ cannot exceed the OOK state (ON/OFF) duration; under ideal synchronization and a $50\%$ duty cycle, optimal $N$ matches this single-state duration. The correlation operation is performed over consecutive blocks $k = 1, \dots, K$ to generate the sequence $\{R_k[l]\}_{k=1}^K$. Specifically, in the $k$-th block, the receiver extracts a received segment $r[m]$ for $m \in [(k-1)N, kN-1]$ and a reference segment $s[m]$ for $m \in [(k-1)N - L_{\max}, kN-1]$. The cross-correlation across delay lags $l \in [0, L_{\max}]$ is evaluated as
\begin{equation}
R_k[l] = \sum_{m=(k-1)N}^{kN-1} r[m] s[m - l], \quad l \in [0, L_{\max}].
\end{equation}

For analytical tractability, we temporarily neglect the sign variations on correlation peaks caused by OOK modulation. Consequently, $R_k[l]$ contains $M$ peaks from the $M$ tags. Since Tag 1 provides the dominant reflection, the receiver estimates its two-way sample delay $2\hat{l}_1$ by locating the global maximum:
\begin{equation}
2\hat{l}_1 = \arg\max_{l \in [0, L_{\max}]} |R_k[l]|.
\end{equation}


Once the delay position $2\hat{l}_1$ of Tag 1 is locked, the receiver continuously samples the subsequent cross-correlation profiles specifically at this designated lag $l=2\hat{l}_1$. This across-block sampling yields a discrete sequence of correlation peaks dedicated to Tag 1, defined as:
\begin{equation}
y_1[k] = R_k[2\hat{l}_1], \quad k = 1, 2, \dots, K.
\end{equation}

Assuming ideal synchronization where correlation windows align with ON/OFF durations, $y_1[k]$ samples the peak amplitude, allowing $\{y_1[k]\}$ to track the state transitions of Tag 1. The receiver then recovers the modulation symbols $\hat{v}_1[k]$ via a threshold test:
\begin{equation}
\hat{v}_1[k] = 
\begin{cases}
1, & |y_1[k]| \ge \gamma_{\text{th}}, \\
0, & |y_1[k]| < \gamma_{\text{th}},
\end{cases}
\end{equation}
where $\gamma_{\text{th}}$ is a predefined threshold. 

Using the estimated delay $2\hat{l}_1$, reconstructed modulation waveform $\hat{v}_1[n]$, and attenuation $\hat{a}_1$, the reflected signal from Tag 1 is reconstructed as
\begin{equation}
\hat{r}_1[n] = \hat{a}_1 s[n - 2\hat{l}_1] \hat{v}_1[n -\hat{l}_1].
\end{equation}

Subtracting $\hat{r}_1[n]$ from the composite signal $r[n]$ cancels this interference:
\begin{equation}
r[n] \leftarrow r[n] - \hat{r}_1[n].
\end{equation}
This residual signal is fed back to the cross-correlation stage. The receiver iteratively extracts the next dominant path and demodulates the remaining tags ($i = 2, \dots, M$) until all tags are recovered.

\subsection{Interference Analysis Under Imperfect SIC}

To understand the statistical variance in the decision metric, we first consider the degrees of freedom (DoF) of the baseband signal $s(t)$. For a signal with bandwidth $B$ and time duration $T$ (which corresponds to the correlation window length $N$ in the discrete domain), its DoF is given by $BT$. An ideal white gaussian noise signal possesses infinite bandwidth and infinite duration, yielding infinite DoF. Its auto-correlation is a Dirac delta function, and the cross-correlation between independent realizations is strictly zero. However, in practical systems, the signal is truncated in both time and frequency. This limited DoF introduces a fluctuation (variance) in the cross-correlation: the lower the DoF, the higher the variance introduced by this truncation. Consequently, in the subsequent derivations, the time-bandwidth product $BT$ serves as a crucial scaling factor for the interference and noise variance. We also introduce two correction factors: $\rho_{\mathrm{peak}}$ $\in (0,1]$ to scale the self-fluctuation variance at the desired auto-correlation peak, and $\rho_{\text{res}}$ $\in (0,1]$ to adjust the residual cross-correlation variance from inter-tag interference. Unlike ideal white gaussian noise, practical OFDM signals contain deterministic components like cyclic prefixes and pilots, which increase correlation and cause the variance to deviate from the $1/BT$ model. As the OFDM signal more closely approximates a Gaussian random process, both $\rho_{\mathrm{peak}}$ and $\rho_{\mathrm{res}}$ converge toward 1.


In a multi-tag scenario, hardware impairments, thermal noise, and synchronization offsets prevent perfect Successive Interference Cancellation (SIC). We thus introduce a residual interference power fraction $\xi \in [0, 1)$ to model the energy leakage from imperfectly cancelled paths. 

When evaluating the decision metric $R_1[2l_i]$ for Tag $i$ over an integration duration $N$, the matched filter output can be explicitly decomposed into four distinct terms:
\begin{equation}
\begin{aligned}
R_1[2l_i] &= \underbrace{\sum_{m=0}^{N-1} a_i s^2[m - 2l_i] v_i[m - l_i]}_{\text{Term 1: Desired Signal of Tag } i} \\
&\quad + \underbrace{\sqrt{\xi} \sum_{m=0}^{N-1} \sum_{j=1}^{i-1} a_j s[m - 2l_j] s[m - 2l_i] v_j[m - l_j]}_{\text{Term 2: Residual Interference from Prior } i-1 \text{ Stronger Tags}} \\
&\quad + \underbrace{\sum_{m=0}^{N-1} \sum_{j=i+1}^{M} a_j s[m - 2l_j] s[m - 2l_i] v_j[m - l_j]}_{\text{Term 3: Uncancelled Interference from Remaining } M-i \text{ Weaker Tags}} \\
&\quad + \underbrace{\sum_{m=0}^{N-1} w[m] s[m - 2l_i]}_{\text{Term 4: Ambient Noise Term}},
\end{aligned}
\label{eq:R_2li_decomposition}
\end{equation}
where the first term represents the target signal (associated with $\rho_{\text{peak}}$). The second term accounts for the multi-access interference (MAI) leaked from the previously processed $i-1$ tags under imperfect cancellation, and the third term denotes the uncancelled MAI from the subsequent $M-i$ weaker tags (both associated with $\rho_{\text{res}}$). The fourth term is the AWGN.

Specifically, based on this decomposition, the total statistical variance of the decision metric comprises three independent components. First, arising from the finite-time energy estimation of the downlink signal, the self-fluctuation term fluctuates around its mean $\mu_i = a_i \sigma_s^2 v_i$, where $\sigma_s^2$ denotes the average power of the signal $s(t)$:
\begin{equation}
\sigma_{\text{self}, i}^2 = \frac{2 \rho_{\text{peak}} a_i^2 \sigma_s^4 v_i^2}{BT}.
\end{equation}
$\rho_{\text{peak}}$ exhibits a special case under normalization. By dividing the cross-correlation output by the local reference energy $\sum_{m=0}^{N-1} s^2[m - 2\hat{l}_1]$, the receiver can eliminate the self-fluctuation of the OFDM signal, driving $\rho_{\text{peak}} \to 0$ in the theoretical model.
Second, since inter-tag delays exceed the correlation mainlobe, cross-terms across distinct tags yield zero mean. Assuming i.i.d. symbols with average power factor $\kappa = \frac{1+\eta^2}{2}$, the total effective MAI variance scaled by $\rho_{\text{res}}$ is given by%
\begin{equation}
\sigma_{\text{MAI}, i}^2 = \frac{\rho_{\text{res}} \kappa \sigma_s^4}{BT} \left( \xi \sum_{j=1}^{i-1} a_j^2 + \sum_{j=i+1}^{M} a_j^2 \right).
\end{equation}%
Finally, due to the statistical independence between $s(t)$ and $w(t)$, their cross-correlation produces zero-mean ambient noise with variance%
\begin{equation}
\sigma_{\text{noise}}^2 = \frac{\sigma_s^2 \sigma_w^2}{BT}.
\end{equation}

\begin{figure*}[!t]
    \centering
    \begin{subfigure}[b]{0.48\textwidth}
        \centering
        \includegraphics[width=\linewidth]{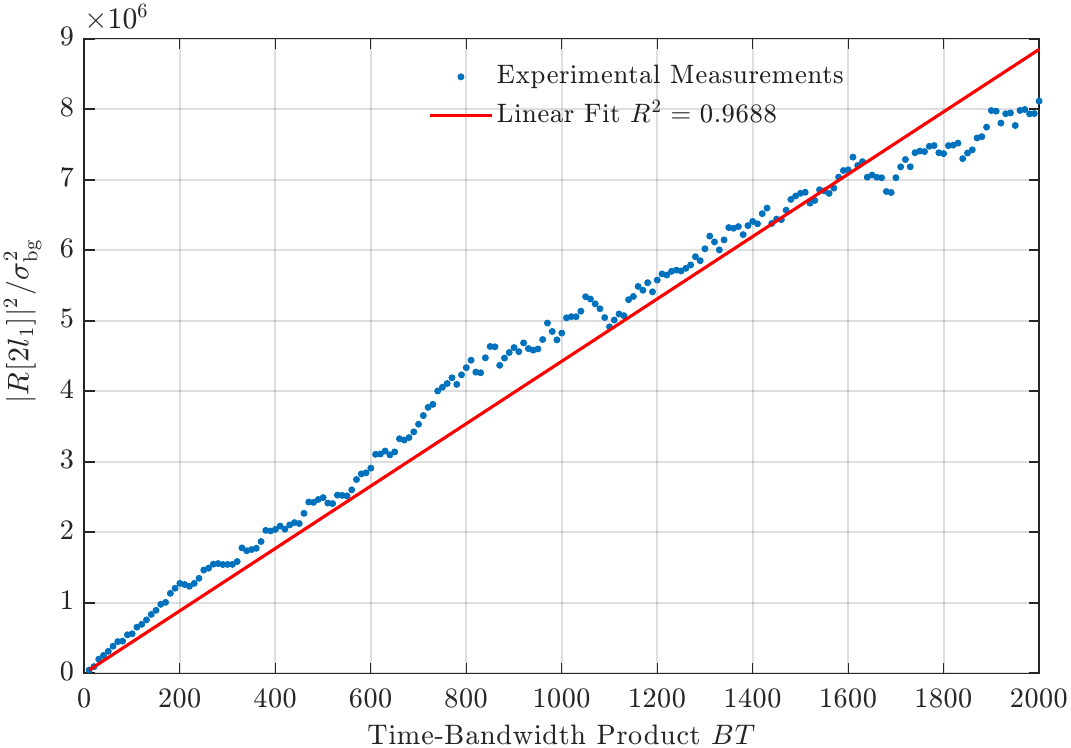}
        \caption{}
        \label{fig:sub1}
    \end{subfigure}
    \hfill
    \begin{subfigure}[b]{0.48\textwidth}
        \centering
        \includegraphics[width=\linewidth]{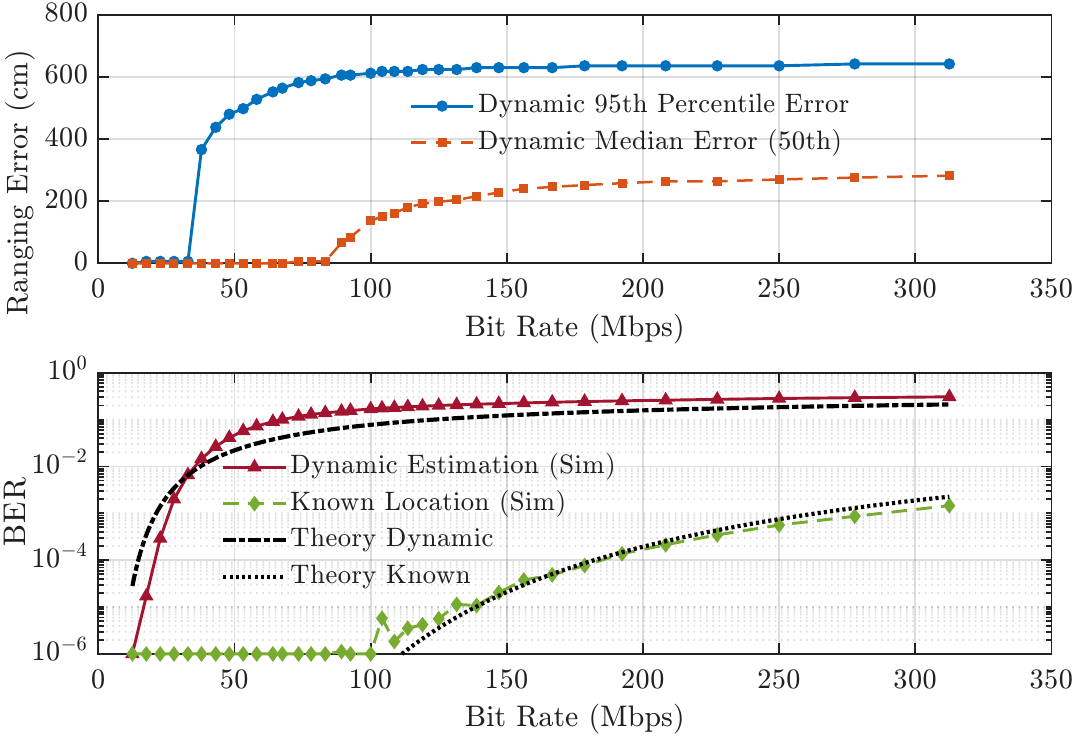}
        \caption{}
        \label{fig:sub2}
    \end{subfigure}
    
    \vspace{0.8em} 
    
    \begin{subfigure}[b]{0.48\textwidth}
        \centering
        \includegraphics[width=\linewidth]{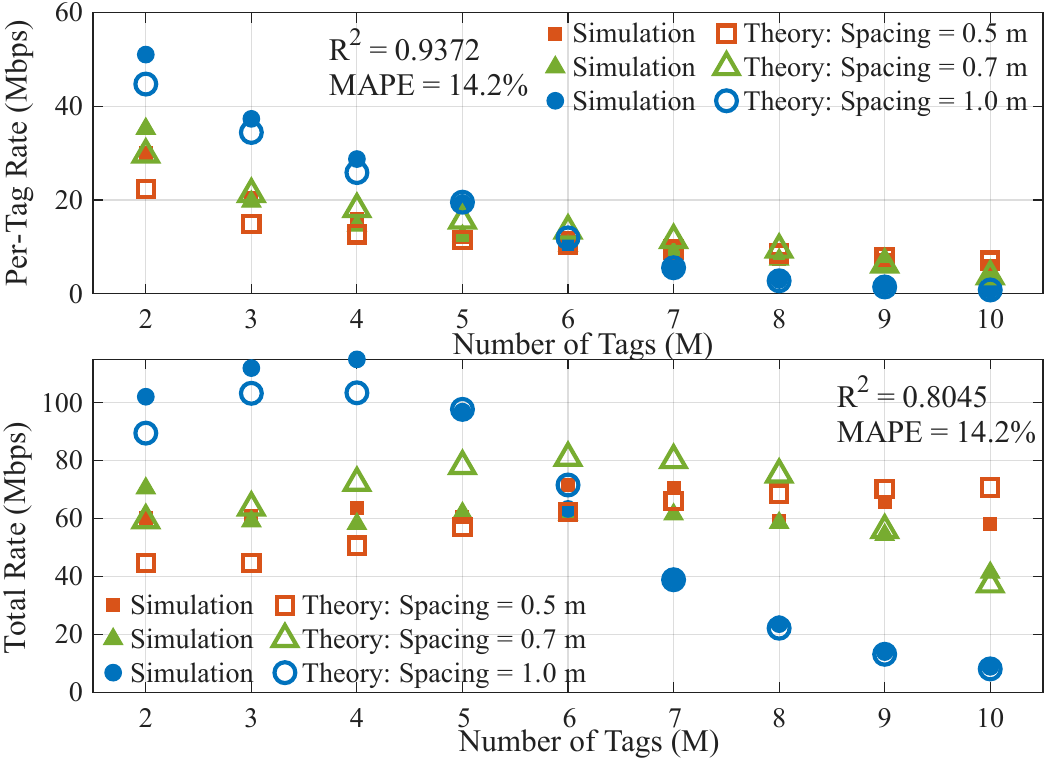}
        \caption{}
        \label{fig:sub3}
    \end{subfigure}
    \hfill
    \begin{subfigure}[b]{0.48\textwidth}
        \centering
        \includegraphics[width=\linewidth]{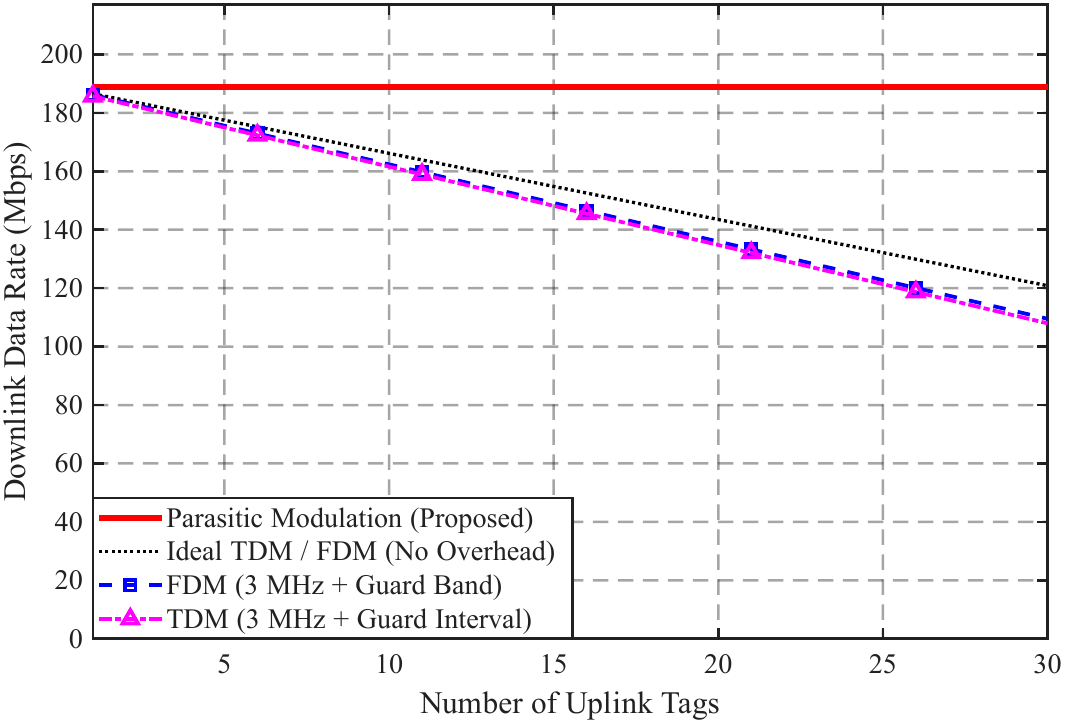}
        \caption{}
        \label{fig:sub4}
    \end{subfigure}
    
    \caption{Performance evaluation: (a) Peak-to-noise metric versus $BT$ product; (b) single-tag ranging error and BER versus bit rate; (c) per-tag rate and sum capacity versus tag count $M$ under different spacings; and (d) downlink data rate versus tag count under TDD/TDM and FDD/FDM benchmarks.}
    \label{fig:entire_experiment}
    \vspace{-6pt}
\end{figure*}

\subsection{Theoretical Bit Error Rate (BER)}
Given a large time-bandwidth product ($BT \gg 1$), the decision metric $y_i[k]$ is approximately Gaussian by the Central Limit Theorem. Hence, demodulating Tag~$i$ reduces to a binary hypothesis testing problem:
\begin{align}
\mathcal{H}_1 \ (\text{ON}):\quad y_i[k] &\sim \mathcal{N}\left(\mu_{1, i}, \sigma_{1, i}^2\right), \\
\mathcal{H}_0 \ (\text{OFF}):\quad y_i[k] &\sim \mathcal{N}\left(\mu_{0, i}, \sigma_{0, i}^2\right).
\end{align}

Under hypothesis $\mathcal{H}_1$ ($v_i = 1$), the peak amplitude is $\mu_{1, i} = a_i \sigma_s^2$, and the total calibrated decision variance is:
\begin{equation}
\sigma_{1, i}^2 = \frac{2 \rho_{\text{peak}} a_i^2 \sigma_s^4}{BT} + \sigma_{\text{MAI}, i}^2 + \frac{\sigma_s^2 \sigma_w^2}{BT}.
\end{equation}

Under hypothesis $\mathcal{H}_0$ ($v_i = \eta$), the mean attenuates to $\mu_{0, i} = \eta a_i \sigma_s^2$ due to residual backscattering, yielding the variance:
\begin{equation}
\sigma_{0, i}^2 = \frac{2 \rho_{\text{peak}} \eta^2 a_i^2 \sigma_s^4}{BT} + \sigma_{\text{MAI}, i}^2 + \frac{\sigma_s^2 \sigma_w^2}{BT}.
\end{equation}

By adopting the widely used quasi-optimal decision threshold $\gamma_{\text{th}} = \frac{\mu_{0, i} \sigma_{1, i} + \mu_{1, i} \sigma_{0, i}}{\sigma_{0, i} + \sigma_{1, i}}$, which equates the tail-error probabilities of both states, the detection performance of Tag $i$ is characterized by the effective $Q$-factor:
\begin{equation}
Q_{\text{factor}, i} = \frac{\mu_{1, i} - \mu_{0, i}}{\sigma_{1, i} + \sigma_{0, i}} = \frac{(1 - \eta) a_i \sigma_s^2}{\sigma_{1, i} + \sigma_{0, i}}.
\end{equation}

Assuming equal prior probabilities for the ON and OFF states, the closed-form analytical BER for Tag $i$ is evaluated via the standard Gaussian $Q$-function:
\begin{equation}
\mathrm{BER}_i \approx Q(Q_{\text{factor}, i}) = \frac{1}{\sqrt{2\pi}} \int_{Q_{\text{factor}, i}}^{+\infty} \exp\left(-\frac{u^2}{2}\right) \mathrm{d}u,
\end{equation}
or equivalently, $\mathrm{BER}_i \approx \frac{1}{2}\operatorname{erfc}\left( \frac{Q_{\text{factor}, i}}{\sqrt{2}} \right)$, where $\operatorname{erfc}(\cdot)$ is the complementary error function.

\section{Performance Evaluation}
To validate the theoretical derivation and evaluate the potential of the proposed RO-ISAC architecture, we conduct both hardware testbed experiments and numerical simulations\footnote{\url{https://github.com/Shao-wireless-lab/Sensing-Based-Communication-A-Non-Orthogonal-Passive-Reflection-Framework-for-ISAC.git}} across single-tag and multi-tag scenarios.

First, we validate the physical assumption using a commercial LED transmitting a 3~MHz Gaussian waveform to a passive retro-reflective tag at 3~m. The light source consists of a $6 \times 34$ LED array (6 parallel branches of 34 LEDs in series) driven by a 102~V DC supply, where the DC bias and arbitrary waveform generator (AWG)-generated AC signals are coupled through a bias tee. The backscattered light from the corner-cube is focused by a central lens with a 12~cm focal length onto a photodiode, whose output is sampled by an oscilloscope. Offline analysis in Fig.~\ref{fig:entire_experiment}(a) demonstrates that the ratio of the squared cross-correlation peak to background noise variance exhibits a linear relationship with the time-bandwidth product $BT$ ($R^2 = 0.9688$), confirming that broader bandwidth and longer integration times improve SINR. 
Due to prototype LED bandwidth limitations, we evaluate the system's wideband potential via simulation with a 250~MHz optical source technically achievable by micro-LEDs\cite{10697462} yet at a higher cost. The signal source transmits 1,024-subcarrier binary phase-shift keying (BPSK)-OFDM signals, where the receiver samples at 2.5~GHz (yielding a distance resolution of 6~cm), and the signal-to-noise ratio (SNR) at 5~m is set to be 15~dB. As shown in Fig.~\ref{fig:entire_experiment}(b), the tag is fixed at 5~m but treated as unknown during demodulation. By dynamically searching the entire region for the cross-correlation peak to estimate the tag's position, the BER and ranging error become coupled, achieving a data rate of 25~Mbps at $\text{BER} \le 10^{-3}$, with theoretical predictions closely matching the empirical simulation results in both trends and numerical values.

Next, we investigate multi-tag access in Fig.~\ref{fig:entire_experiment}(c) across three spatial spacings ($\Delta d \in \{0.5, 0.7, 1.0\}$~m, with the closest tag at 5~m). All tags are configured with identical data rates and operate under a target threshold of $\text{BER} \le 10^{-3}$ with fixed locations.
Here, the per-tag throughput is bounded by the bottleneck tag suffering the lowest SINR. At small tag counts ($M \le 4$), $\Delta d = 1.0$~m achieves the highest rate because the bottleneck resides at Tag~1, and wider spacing mitigates interference from subsequent tags. As $M$ grows beyond 7, severe $d^4$ geometric path loss shifts the bottleneck to remote edge tags, causing their per-tag rate to drop below denser deployments. This shift in the performance bottleneck is also accurately captured and substantiated by our theoretical analytical model. Furthermore, the theoretical analytical model demonstrates strong agreement with the simulation results across the data-rate range of 1--50~Mbps, yielding an \(R^2\) of 0.9372. The \(R^2\) indicates that the closed-form model captures the overall trend of the simulation results, while maintaining an average mean absolute percentage error (MAPE) of 14.02\% over this broad range.

Finally, Fig.~\ref{fig:entire_experiment}(d) compares our parasitic modulation against conventional TDD/TDM and FDD/FDM duplexing and multiplexing benchmarks. In traditional retro-reflective ISAC architectures, ranging relies on reflected echoes tied directly to downlink illumination. To prevent these retro-reflected echoes from colliding with active uplink signals at the receiver, conventional systems enforce strict orthogonal slicing between downlink and uplink via TDD or FDD, as well as TDM or FDM among multiple uplink tags. Consequently, as the number of uplink tags increases, with each tag requiring a 3~Mbps data rate, the additional uplink traffic directly consumes downlink resources, resulting in a linear decline in downlink throughput that is further aggravated by guard bands and guard intervals. In contrast, our parasitic mechanism simultaneously embeds both uplink transmission and retro-reflective ranging into the existing downlink waveform, and achieves true zero spectral overhead.


\section{Conclusion}
In this paper, we tackled the severe downlink throughput degradation in conventional RO-ISAC systems caused by orthogonal resource slicing between ranging echoes and uplink communications. To eliminate this bottleneck, we proposed a parasitic mechanism that embeds active uplink transmissions and retro-reflective ranging directly into the downlink waveform. Both theoretical and numerical results demonstrate that our design achieves multi-tag uplink access without degrading downlink capacity, while preserving tag simplicity. These characteristics make the proposed approach promising for scenarios where preserving the performance of an existing downlink is critical, as well as where orthogonal resource multiplexing is difficult or impractical. Importantly, this parasitic paradigm generalizes across different wave domains. While light requires hundreds of megahertz to separate targets, sound needs merely tens of kilohertz. This provides practical insights for broader applications.

\bibliographystyle{IEEEtran}
\bibliography{reference.bib}
\end{document}